\documentclass[letterpaper]{article} % DO NOT CHANGE THIS
\usepackage{aaai25}  % DO NOT CHANGE THIS
\usepackage{times}  % DO NOT CHANGE THIS
\usepackage{helvet}  % DO NOT CHANGE THIS
\usepackage{courier}  % DO NOT CHANGE THIS
\usepackage[hyphens]{url}  % DO NOT CHANGE THIS
\usepackage{graphicx} % DO NOT CHANGE THIS
\usepackage{natbib}  % DO NOT CHANGE THIS AND DO NOT ADD ANY OPTIONS TO IT
\usepackage{caption} % DO NOT CHANGE THIS AND DO NOT ADD ANY OPTIONS TO IT
\usepackage{amssymb}
\usepackage{multirow}
\usepackage{bbding}
\usepackage{pifont}

\usepackage{algorithm}
\usepackage{algorithmic}

\usepackage{newfloat}
\usepackage{listings}
\DeclareCaptionStyle{ruled}{labelfont=normalfont,labelsep=colon,strut=off} % DO NOT CHANGE THIS
\floatstyle{ruled}
\newfloat{listing}{tb}{lst}{}
\floatname{listing}{Listing}
\title{SPU-IMR: Self-supervised Arbitrary-scale Point Cloud Upsampling via Iterative Mask-recovery Network}
\author{
Ziming Nie\textsuperscript{\rm 1}, Qiao Wu\textsuperscript{\rm 1}, Chenlei Lv\textsuperscript{\rm 2}, Siwen Quan\textsuperscript{\rm 3}, Zhaoshuai Qi\textsuperscript{\rm 1}, Muze Wang\textsuperscript{\rm 1}, Jiaqi Yang\textsuperscript{\rm 1}*
}
\affiliations{
    \textsuperscript{\rm 1}Northwestern Polytechnical University, Xi'an, China\\
    \textsuperscript{\rm 2}Shenzhen University, Shenzhen, China\\
    \textsuperscript{\rm 3}Chang'an University, Xi'an, China\\
    \{nzm, qiaowu\}@mail.nwpu.edu.cn, chenleilv@mail.bnu.edu.cn, jqyang@nwpu.edu.cn
}

\usepackage{bibentry}
\begin{document}

\maketitle

\begin{abstract}
Point cloud upsampling aims to generate dense and uniformly distributed point sets from sparse point clouds. Existing point cloud upsampling methods typically approach the task as an interpolation problem. They achieve upsampling by performing local interpolation between point clouds or in the feature space, then regressing the interpolated points to appropriate positions. By contrast, our proposed method treats point cloud upsampling as a global shape completion problem. Specifically, our method first divides the point cloud into multiple patches. Then, a masking operation is applied to remove some patches, leaving visible point cloud patches. Finally, our custom-designed neural network iterative completes the missing sections of the point cloud through the visible parts. During testing, by selecting different mask sequences, we can restore various complete patches. A sufficiently dense upsampled point cloud can be obtained by merging all the completed patches. We demonstrate the superior performance of our method through both quantitative and qualitative experiments, showing overall superiority against both existing self-supervised and supervised methods. 
%The code is available at https://github.com/hapifuzi/spu-imr.

\end{abstract}

\begin{links}
  \link{Code}{https://github.com/hapifuzi/spu-imr}
\end{links}

\section{Introduction}
\label{sec:intro}
In recent years, as an important 3D representation, point clouds have attracted attention from both academia and industry \cite{qian2021deep, yang2023mutual, cheng2023sampling, zhang20233d, du2024arbitrary}. With the advancement of technology, high-quality point clouds are increasingly in demand for downstream tasks such as autonomous driving \cite{zeng2018rt3d, li2020deep}, virtual reality \cite{blanc2020genuage, zhang2024plug}, and 3D reconstruction \cite{park2019deepsdf, mescheder2019occupancy}. The raw point clouds are primarily obtained through 3D scanning devices. However, due to the inherent limitations of these devices, the exported point clouds are often sparse and non-uniform. Various learning-based approaches have been proposed to tackle the task of generating dense and uniformly distributed point clouds from the sparse point cloud inputs.

\begin{figure}[t]
    \centering
    \includegraphics[width=1.0\linewidth]{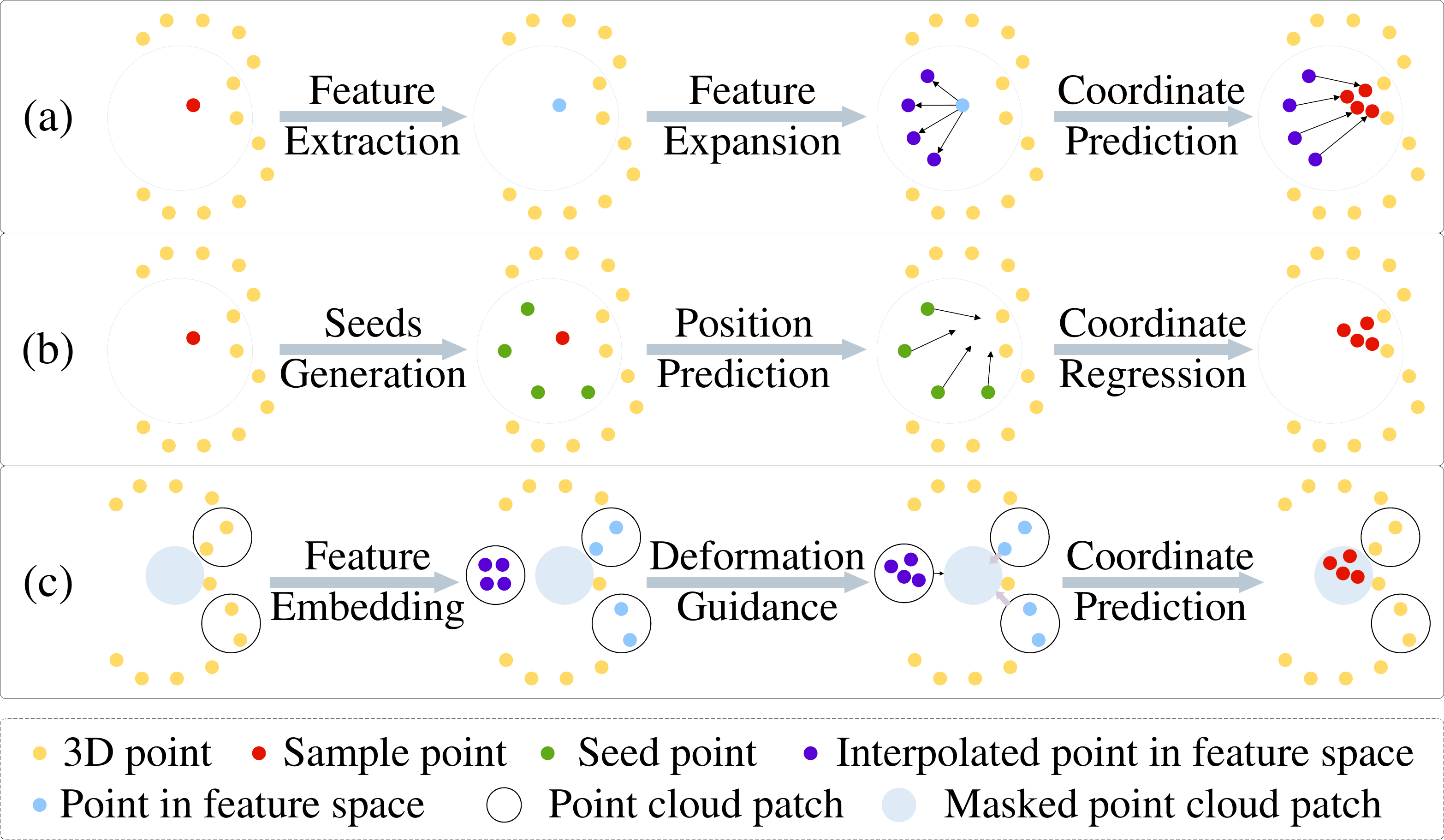}
    \caption{Comparative illustrations of mainstream methods and our method in 2D. Take a point or a patch as an example: (a) Feature-space-based interpolation methods interpolate points in feature space. (b) Point-cloud-based interpolation methods directly interpolate points around the input. (c) Our proposed method predicts the masked point cloud patch's local shape and topological changes by leveraging the visible patches.
}
    \label{fig:fig_0}
\end{figure}

Existing point cloud upsampling methods can be broadly categorized into two locally interpolation-based approaches: feature-space-based interpolation methods (Fig. \ref{fig:fig_0} (a)) and point-cloud-based interpolation methods (Fig. \ref{fig:fig_0} (b)). Feature-space-based interpolation typically involves feature extraction, feature expansion, and coordinate prediction. First, these methods extract features from sparse point clouds. Then, they expand the features of each point in the feature space. In particular, interpolated points are generated within the feature space. Finally, they regress the expanded features into a 3D form through coordinate prediction. However, directly predicting 3D point cloud coordinates after feature expansion is challenging, resulting in limited accuracy. By contrast, point-cloud-based interpolation methods insert rough points (seeds) directly around the input point clouds and then predict the offsets of each seed through a designed neural network. These methods often require generating many seed points and are prone to generating outliers due to errors in neural networks. While most interpolation-based methods can learn point clouds' local and global features, they face challenges in capturing variations and connections within the local topological structures. Additionally, most existing methods require dense point clouds for supervised training, which is challenging to apply in real-world scenarios.

To address these issues, inspired by Point-MAE \cite{pang2022masked}, we propose an iterative mask-recovery network in a self-supervised manner for point cloud upsampling, called SPU-IMR, which models upsampling as a completion problem. As shown in Fig. \ref{fig:fig_0}, unlike existing interpolation-based methods, our approach predicts the local shape and topological changes of missing point cloud patches based on known topological structures. Specifically, our method proceeds as follows. First, we select certain points as centers and cluster the surrounding points into patches centered on them. Next, some point cloud patches are masked according to a given mask sequence. Subsequently, learnable point cloud patches are randomly generated. Guided by the position encoding of the center points and the features of the visible point cloud patches, we iteratively deform the generated patches to restore the topological structure characteristics of the masked patches. Therefore, the masked patches can be completed by generated patches. Through training and learning with our designed neural network, input point clouds can complete the masked portions of the point cloud patches using this mask-recovery operation.

To achieve point cloud upsampling at an arbitrary scale, we propose a multi-mask-recovery (MMR) module during the testing phase. First, different mask sequences are selected to create various combinations of visible patches. Then, visible patch combinations are completed to obtain various output patch combinations. When the output patch combinations cover all point cloud patches and are sufficient enough, we combine them to obtain a dense point cloud. Finally, we can get any desired ratio of upsampled point clouds through simple outlier removal and farthest point sampling (FPS). Our end-to-end training approach, combined with iterative adjustments for the needed completion patches, ensures that the generated results are stable with a low probability of noise generation. Additionally, sufficient dense point clouds allow for a relatively large spacing between point clouds when using FPS, resulting in uniform outcomes. Our contributions are summarized as follows:
\begin{itemize}
    \item To the best of our knowledge, we are the first to adopt an explicit completion approach to achieve point cloud upsampling.
    \item We propose a novel self-supervised-based framework SPU-IMR to achieve point cloud upsampling. SPU-IMR is an iterative mask-recovery framework, which can accurately predict the local topological structure.
    \item At testing time, we propose the MMR module to achieve arbitrary-scale point cloud upsampling and generate uniform upsampled point clouds.
\end{itemize}

\begin{figure*}[t]
    \centering
    \includegraphics[width=0.92\linewidth]{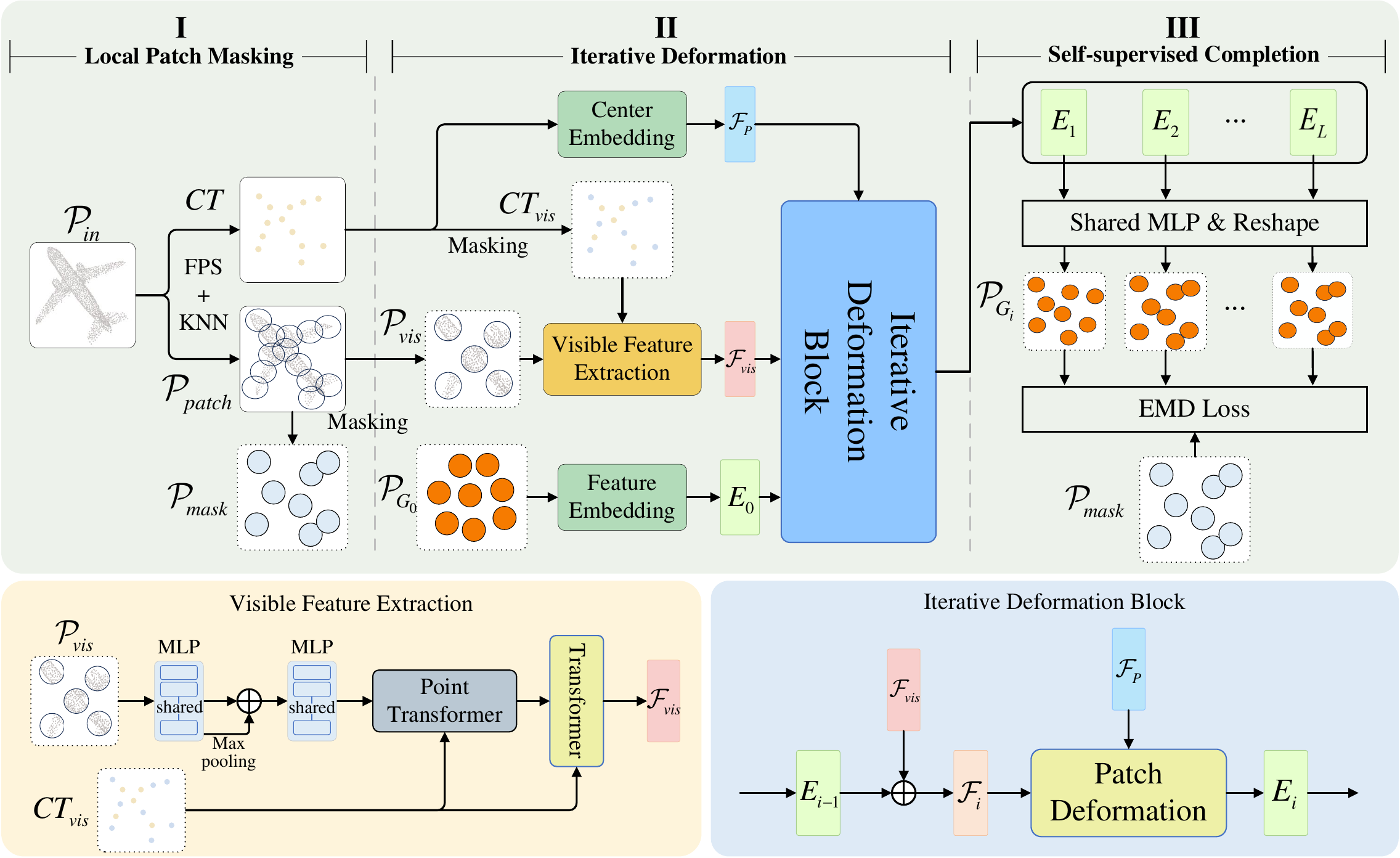}
    \caption{Illustration of the proposed SPU-IMR method. First, we obtain the center points and point cloud patches using FPS and KNN. We then remove some point cloud patches based on a given mask ratio, leaving only the visible patches. Guided by encoding the visible patch features and the center point position embedding, learnable generated point cloud patches are iteratively transformed and optimized in the iterative deformation block module, completing the removed point cloud patches. Finally, we use the EMD between all output point cloud patches and the real point cloud patches as the loss function.
}
    \label{fig:fig_1}
\end{figure*}

\begin{figure*}[t]
    \centering
    \includegraphics[width=0.9\linewidth]{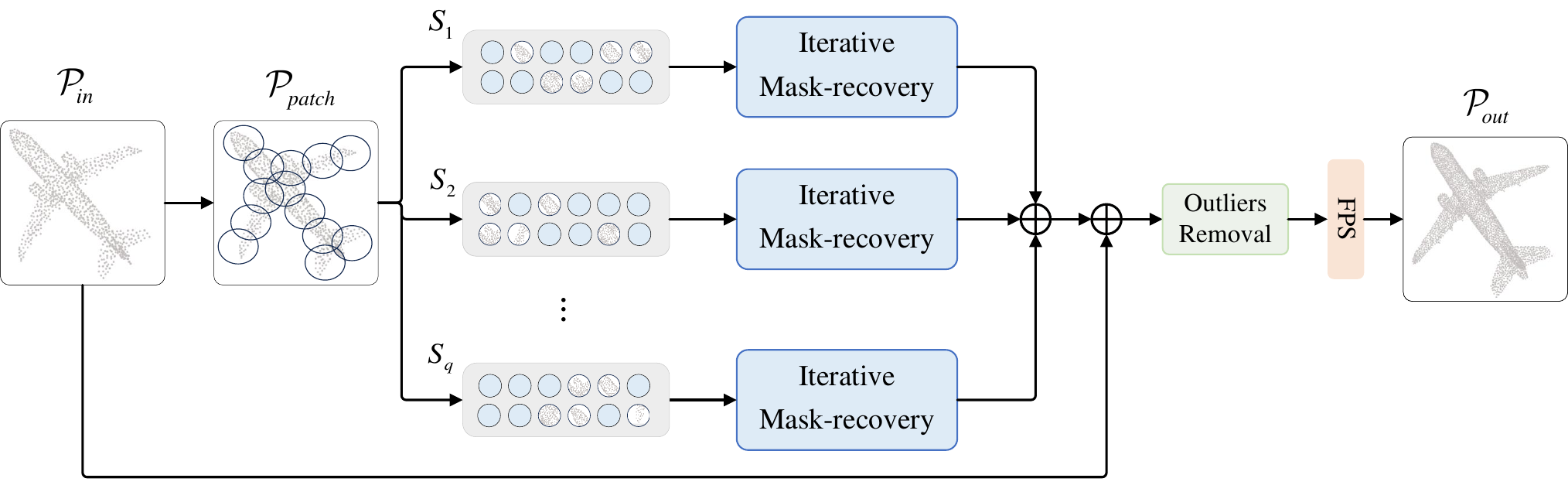}
    \caption{Illustration of the proposed multi-mask-recovery module. We first divide the point cloud into patches and then mask them using the selected mask sequence. Afterward, we import these patches into the Iterative Mask-recovery module. All output results will be merged with $\mathcal{P}_{in}$. Finally, after removing outliers through the Outliers Removal module, the desired upsampling point cloud can be obtained through FPS.
}
    \label{fig:fig_2}
\end{figure*}

\section{Related Work}

\subsection{Feature-space-based Interpolation Methods}
Feature-space-based interpolation methods typically divide the point cloud upsampling network into three crucial steps: feature extraction, feature expansion, and coordinate prediction. As a pioneer work, PU-Net \cite{yu2018pu} first proposes this three-step framework to effectively learn the mapping from sparse point clouds to dense point clouds. Subsequently, with advancements in network designs, more sophisticated end-to-end networks have emerged \cite{yu2018ec, yifan2019patch, li2019pu, qian2020pugeo, zhao2021sspu, ye2021meta, Mao2022pu}. To better encode local point information from point neighborhoods, PU-GCN \cite{qian2021pu} incorporates graph convolutional networks to encode point cloud features. The transformer structure is first introduced by PU-Transformer \cite{qiu2022pu} for extracting fine-grained point cloud features. SPU-Net \cite{liu2022spu} introduces self-supervised learning by innovatively designing a coarse-to-fine structure for point cloud upsampling. Drawing inspiration from the Dis-PU \cite{li2021point} framework, SSPU-Net \cite{wang2023sspu} integrates frequency-aware attention mechanisms to extract edge and contour information from point clouds. However, these methods directly predict point cloud coordinates after the feature expansion step, which is challenging work and usually results in lower accuracy. Our proposed method captures local topological connections and variations between point cloud patches, enabling the generation of highly accurate upsampled point clouds.

\subsection{Point-cloud-based Interpolation Methods}
Point-cloud-based interpolation methods first insert rough point clouds directly between the input points. Then, they predict the correct location of rough points through the learned shape and structural features \cite{lv2021approximate, lv2022intrinsic, wei2023ipunet, rong2024repkpu, liu2024pu}. As a pioneering work in shape expression, NePs \cite{feng2022neural} is the first to introduce implicit neural fields for representing the global shapes of point clouds. SAPCU \cite{zhao2022self} and PUSS-AS \cite{zhao2023self} propose self-supervised approaches by regressing seed points to the object surface using learned signed distance functions. P2PNet is proposed by Grad-PU \cite{he2023grad} to refine seed points through learned point-to-point distance functions. To further remove holes, IFLDI \cite{li2023learning} guides seed point projection by learning unsigned distance fields and local distance indicators. More recently, SPU-PMD \cite{liu2024spu} designs a series of mesh deformers to improve coarse points generated from the mesh interpolator. However, these methods are easily making incorrect predictions for rough points. Our proposed method avoids generating rough points and can produce arbitrary-scale upsampled point clouds with just one-time training.

\section{The Proposed Method}

\subsection{Overview}
We propose a novel iterative mask-recovery framework based on self-supervised learning, which can upsample point clouds at arbitrary scales through one-time training. The overview of our proposed method is shown in Fig. \ref{fig:fig_1}. First, we divide the input point cloud into multiple point cloud patches and mask some of them through a mask sequence. Second, learnable point cloud patches are randomly generated. The position encoding of the center points and the features of the visible patches are used to iteratively guide the deformation of the learnable patches. Finally, we achieve arbitrary-scale point cloud upsampling through our proposed MMR module.

\subsection{Local Patch Masking}
Given a sparse point cloud $\mathcal{P}_{in}=\{p_i\}^N_{i=1},p_i\in \mathbb{R}^{3}$, we random select center points $CT\in \mathbb{R}^{n \times3}$ through FPS algorithm. Then, the K-Nearest Neighbors (KNN) algorithm is applied on the center points to cluster $k$ nearest points and construct point cloud patches: 

\begin{equation}
    \mathcal{P}_{patch} = \{ {\mathcal{P}_1},{\mathcal{P}_2},{\mathcal{P}_3},...,{\mathcal{P}_n}\} ,{\mathcal{P}_i} \in {\mathbb{R}^{k \times 3}}\,,
    \label{formula_1}
\end{equation}
where ${\mathcal{P}_i}$ are point cloud patches and $\mathcal{P}_{patch} \in \mathbb{R}{^{n \times k \times 3}}$.

We select mask ratio $m$ to obtain a mask sequence of length $n$. We randomly mask the point cloud patches through mask sequence and obtain visible patches $\mathcal{P}_{vis}\in {\mathbb{R}^{(1-m)n \times k \times 3}} $ and masked patches $\mathcal{P}_{mask}\in {\mathbb{R}^{mn \times k \times 3}} $.

Finally, we transform the neighboring points within each patch into a local coordinate system centered at $CT$ to standardize the coordinates. Additionally, we select sufficient center points to create overlapping point cloud patches, leading to a better understanding of local topological changes.

\subsection{Iterative Deformation for Recovery}
\subsubsection{Visible Feature Extraction.}
To fully explore the features of visible point cloud patches, we design a Visible Feature Extraction encoder. As shown in Fig. \ref{fig:fig_1} of Visible Feature Extraction, we first use shared Multi-Layer-Perceptron (MLP) and max-pooling layers to extract local and global features of visible patches $\mathcal{F}_{m} \in {\mathbb{R}^{(1-m)n \times C}}$. Next, the Point Transformer layer \cite{zhao2021point} $PointTr$ is applied to mine the contextual features of visible point cloud patches, aiding in understanding the overall topological structure:

\begin{equation}
    \mathcal{F}_{v} = PointTr(\mathcal{F}_{m}, CT_{vis}),\mathcal{F}_{v} \in {\mathbb{R}^{(1-m)n \times C}}\,,
    \label{formula_2}
\end{equation}
where $CT_{vis}\in {\mathbb{R}^{(1-m)n \times 3}}$ is visible part of $CT$.

Finally, we import the embedded $CT_{vis}$ and $\mathcal{F}_{v}$ into the Transformer module $Tr$ for visible feature enhancement:

\begin{equation}
    \mathcal{F}_{vis} = Tr(\mathcal{F}_{v}, Embed(CT_{vis})), \mathcal{F}_{vis}\in {\mathbb{R}^{(1-m)n \times C}}\,.
    \label{formula_3}
\end{equation}

\subsubsection{Iterative Deformation Block.}
We design the iterative deformation block to iteratively deform the generated learnable point cloud patches $\mathcal{P}_{G_0}\in {\mathbb{R}^{mn \times k \times 3}} $, ensuring they accurately restore the topological structure of the masked point cloud patches $\mathcal{P}_{mask}\in {\mathbb{R}^{mn \times k \times 3}} $. Initially, we embed the positional information of the center point to obtain feature $\mathcal{F}_{P}\in {\mathbb{R}^{n \times C}} $. Then we embed the generated point cloud patches to obtain embedded feature ${E}_{0}\in {\mathbb{R}^{mn \times C}} $. Subsequently, the extracted visible feature $\mathcal{F}$ and positional encoding $\mathcal{F}_{P}$ will guide the iterative deformation of the embedded feature ${E}_{0}$.

As shown in Fig. \ref{fig:fig_1} of iterative deformation block with $L$ layers. For the $i \in [1,2,..., L]$ layer, the last layer patch embedding $E_{i-1}$ is served as the input. First, ${E}_{i-1}$ and $\mathcal{F}_{vis}$ are concatenated to obtain the feature $\mathcal{F}_{i}\in \mathbb{R}^{n \times C}$. Next, $\mathcal{F}_{i}$ along with the position encoding $\mathcal{F}_{P}$, are fed into the Patch Deformation module $Deform$ to produce the deformed patch embedding ${E}_{i}\in {\mathbb{R}^{mn \times C}}$. The Patch Deformation module is composed of multiple transformers, allowing visible patches to efficiently guide the deformation of generated patches within the feature space. Finally, after passing through a LayerNorm layer, we retain only the features of the generated patches to the next layer. The overall process can be expressed as follows:

\begin{equation}
    E_{0} = Embed(\mathcal{P}_{G_0})\,,
    \label{formula_4}
\end{equation}

\begin{equation}
    \mathcal{F}_{i} = Concat(E_{i-1}, \mathcal{F}_{vis})\,,
    \label{formula_5}
\end{equation}

\begin{equation}
    E_{i} = Deform(\mathcal{F}_{i}, \mathcal{F}_{P}), 1\leq i\leq L \,,
    \label{formula_6}
\end{equation}
where $L$ is the number of layers in the iterative deformation block. 

\subsection{Self-supervised Completion for Upsampling}

As shown in Fig. \ref{fig:fig_1}, a deformed point cloud is generated in each layer through the shared coordinate prediction module $SCP$, which includes a single MLP followed by a reshape operation. Specifically, 
On each embedding feature ${E}_{i}$, we will output the deformed point cloud $\mathcal{P}_{G_i}\in {\mathbb{R}^{mn \times k \times 3}}$: 

\begin{equation}
    \mathcal{P}_{G_i} = SCP({E}_{i})\,, 1\leq i\leq L \,.
    \label{formula_7}
\end{equation}

We apply the Earth Mover's Distance (EMD) as the loss function. EMD measures the minimum cost of turning one of the point sets into the other, which is defined as:

\begin{equation}
    \mathcal{L}{(P, Q)} = \mathop {\min }\limits_{\phi :{P} \to \mathcal{Q}} \sum\limits_{x \in {P}} {{{\left\| {x - \phi \left( x \right)} \right\|}_2}}\,,
    \label{formula_8}
\end{equation}
where $P$ and $Q$ represent the deformed point cloud patches $\mathcal{P}_{G_i}$ and the corresponding ground truth $\mathcal{P}_{mask}$.

Compared to the commonly used Chamfer Distance loss function, EMD loss helps prevent the deformed patches from clustering, thereby reducing the risk of model overfitting. Additionally, EMD enables the deformed patches to align more closely with the topological structure of the masked patches. The final loss function is defined as:
\begin{equation}
    \mathcal{L} = \sum\limits_{i = 1}^l {{\mathcal{L}}{(\mathcal{P}_{G_i}, \mathcal{P}_{mask})}} \,.
    \label{formula_9}
\end{equation}

\begin{table*}[t]
\centering
\setlength{\tabcolsep}{0.78mm}
\small
\begin{tabular}{cccc|ccccc|ccccc|ccccc}

\hline
\multicolumn{4}{c|}{Upsampling Ratio}                                                                                                         & \multicolumn{5}{c|}{2$\times$}                                                                                                                                                                                 & \multicolumn{5}{c|}{4$\times$}                                                                                                                                                                                 & \multicolumn{5}{c}{8$\times$}                                                                                                                                                                                  \\ \hline
\multicolumn{3}{c|}{Metric($10^{-3}$)}                                                                                         & GT           & \multicolumn{1}{c|}{CD$\downarrow$}         & \multicolumn{1}{c|}{EMD$\downarrow$}        & \multicolumn{1}{c|}{HD$\downarrow$}         & \multicolumn{1}{c|}{F-Score$\uparrow$}      & P2F$\downarrow$        & \multicolumn{1}{c|}{CD$\downarrow$}         & \multicolumn{1}{c|}{EMD$\downarrow$}        & \multicolumn{1}{c|}{HD$\downarrow$}         & \multicolumn{1}{c|}{F-Score$\uparrow$}      & P2F$\downarrow$        & \multicolumn{1}{c|}{CD$\downarrow$}         & \multicolumn{1}{c|}{EMD$\downarrow$}        & \multicolumn{1}{c|}{HD$\downarrow$}         & \multicolumn{1}{c|}{F-Score$\uparrow$}      & P2F$\downarrow$        \\ \hline
\multicolumn{1}{c|}{\multirow{3}{*}{\begin{tabular}[c]{@{}c@{}}Fixed\\ Ratio\end{tabular}}}     & \multicolumn{2}{c|}{PU-Net}  & \Checkmark & \multicolumn{1}{c|}{\bf{16.1}}                   & \multicolumn{1}{c|}{28.1}                   & \multicolumn{1}{c|}{26.3}                   & \multicolumn{1}{c|}{48.2}                   & \bf7.63                   & \multicolumn{1}{c|}{16.6}                   & \multicolumn{1}{c|}{23.3}                   & \multicolumn{1}{c|}{23.8}                   & \multicolumn{1}{c|}{48.0}                   & 7.97                   & \multicolumn{1}{c|}{15.3}                   & \multicolumn{1}{c|}{34.8}                   & \multicolumn{1}{c|}{30.3}                   & \multicolumn{1}{c|}{48.3}                   & 8.13                   \\ \cline{2-19} 
\multicolumn{1}{c|}{}                                                                           & \multicolumn{2}{c|}{PU-GAN}  & \Checkmark & \multicolumn{1}{c|}{19.9}                   & \multicolumn{1}{c|}{29.6}                   & \multicolumn{1}{c|}{27.6}                   & \multicolumn{1}{c|}{\underline{48.7}}                   & 9.59                   & \multicolumn{1}{c|}{15.1}                   & \multicolumn{1}{c|}{\bf20.4}                   & \multicolumn{1}{c|}{23.3}                   & \multicolumn{1}{c|}{49.0}                   & 7.48                   & \multicolumn{1}{c|}{13.3}                   & \multicolumn{1}{c|}{\underline{20.1}}                   & \multicolumn{1}{c|}{24.5}                   & \multicolumn{1}{c|}{48.9}                   & 6.55                   \\ \cline{2-19} 
\multicolumn{1}{c|}{}                                                                           & \multicolumn{2}{c|}{PU-GCN}  & \Checkmark & \multicolumn{1}{c|}{\underline{16.4}}                   & \multicolumn{1}{c|}{27.5}                   & \multicolumn{1}{c|}{\underline{25.4}}                   & \multicolumn{1}{c|}{48.6}                   & \underline{7.74}                   & \multicolumn{1}{c|}{15.0}                   & \multicolumn{1}{c|}{\underline{21.0}}                   & \multicolumn{1}{c|}{\bf22.1}                   & \multicolumn{1}{c|}{48.7}                   & 7.23                   & \multicolumn{1}{c|}{11.7}                   & \multicolumn{1}{c|}{\bf18.8}                   & \multicolumn{1}{c|}{\bf20.7}                   & \multicolumn{1}{c|}{48.4}                   & 5.68                   \\ \hline
\multicolumn{1}{c|}{\multirow{4}{*}{\begin{tabular}[c]{@{}c@{}}Arbitrary\\ Ratio\end{tabular}}} & \multicolumn{2}{c|}{Grad-PU} & \Checkmark & \multicolumn{1}{c|}{17.2}                   & \multicolumn{1}{c|}{\underline{25.9}}                   & \multicolumn{1}{c|}{41.9}                   & \multicolumn{1}{c|}{\underline{48.7}}                   & 8.36                   & \multicolumn{1}{c|}{\bf13.8}                   & \multicolumn{1}{c|}{24.3}                   & \multicolumn{1}{c|}{36.2}                   & \multicolumn{1}{c|}{\bf49.5}                   & \underline{7.19}                   & \multicolumn{1}{c|}{13.4}                   & \multicolumn{1}{c|}{21.8}                   & \multicolumn{1}{c|}{40.1}                   & \multicolumn{1}{c|}{\underline{49.1}}                   & 7.48                   \\ \cline{2-19} 
\multicolumn{1}{c|}{}                                                                           & \multicolumn{2}{c|}{SAPCU}   & \XSolidBrush & \multicolumn{1}{c|}{20.3}                   & \multicolumn{1}{c|}{26.2}                   & \multicolumn{1}{c|}{43.2}                   & \multicolumn{1}{c|}{47.5}                   & 10.3                   & \multicolumn{1}{c|}{14.7}                   & \multicolumn{1}{c|}{23.6}                   & \multicolumn{1}{c|}{42.0}                   & \multicolumn{1}{c|}{47.4}                   & 7.34                   & \multicolumn{1}{c|}{\underline{11.0}}                   & \multicolumn{1}{c|}{24.5}                   & \multicolumn{1}{c|}{41.8}                   & \multicolumn{1}{c|}{47.4}                   & 5.49                   \\ \cline{2-19} 
\multicolumn{1}{c|}{}                                                                           & \multicolumn{2}{c|}{PUSS-AS} & \XSolidBrush & \multicolumn{1}{c|}{20.1}                   & \multicolumn{1}{c|}{\bf24.3}                   & \multicolumn{1}{c|}{32.2}                   & \multicolumn{1}{c|}{47.7}                   & 10.1                   & \multicolumn{1}{c|}{14.6}                   & \multicolumn{1}{c|}{22.1}                   & \multicolumn{1}{c|}{30.8}                   & \multicolumn{1}{c|}{47.6}                   & 7.22                   & \multicolumn{1}{c|}{\bf10.9}                   & \multicolumn{1}{c|}{23.4}                   & \multicolumn{1}{c|}{30.5}                   & \multicolumn{1}{c|}{47.4}                   & \underline{5.40}                   \\ \cline{2-19} 
\multicolumn{1}{c|}{}                                                                           & \multicolumn{2}{c|}{Ours}    & \XSolidBrush & \multicolumn{1}{c|}{18.9} & \multicolumn{1}{c|}{\bf24.3} & \multicolumn{1}{c|}{\bf25.2} & \multicolumn{1}{c|}{\bf49.1} & 9.48 & \multicolumn{1}{c|}{\underline{13.9}} & \multicolumn{1}{c|}{22.2} & \multicolumn{1}{c|}{\underline{22.7}} & \multicolumn{1}{c|}{\underline{49.2}} & \bf6.85 & \multicolumn{1}{c|}{11.1} & \multicolumn{1}{c|}{24.3} & \multicolumn{1}{c|}{\underline{22.3}} & \multicolumn{1}{c|}{\bf49.2} & \bf5.31 \\ \hline
\multicolumn{4}{c|}{Ranking(S/A/AL)}                                                                                            & \multicolumn{1}{c|}{1/2/4}                  & \multicolumn{1}{c|}{1/1/1}                  & \multicolumn{1}{c|}{1/1/1}                  & \multicolumn{1}{c|}{1/1/1}                  & 1/2/4                  & \multicolumn{1}{c|}{1/2/2}                  & \multicolumn{1}{c|}{2/2/4}                  & \multicolumn{1}{c|}{1/1/2}                  & \multicolumn{1}{c|}{1/2/2}                  & 1/1/1                  & \multicolumn{1}{c|}{3/3/3}                  & \multicolumn{1}{c|}{2/3/5}                  & \multicolumn{1}{c|}{1/1/2}                  & \multicolumn{1}{c|}{1/1/1}                  & 1/1/1                  \\ \hline
\end{tabular}
%}
\caption{Performance comparison with SoTA methods regarding CD, EMD, HD, F-Score, and P2F. `\Checkmark' denotes fully-supervised methods; `\XSolidBrush' denotes self-supervised methods. Ranking (S/A/AL) represents the rankings of our proposed method in self-supervised, arbitrary-scale upsampling, and all methods. Bold and underline denote the best and second-best methods}
\label{tab:tab_1}
\end{table*}

\begin{table*}[t]
\centering
\setlength{\tabcolsep}{1.6mm}
\small
\begin{tabular}{cccc|cccc|cccc|cccc}
\hline
\multicolumn{4}{c|}{Upsampling Ratio}                                                                                                                        & \multicolumn{4}{c|}{2$\times$}                                                                                                                                                                               & \multicolumn{4}{c|}{4$\times$}                                                                                                                                                                               & \multicolumn{4}{c}{8$\times$}                                                                                                                                                                                \\ \hline
\multicolumn{3}{c|}{$p$($10^{-2}$)}                                                                                            & \multicolumn{1}{c|}{GT}     & \multicolumn{1}{c|}{0.2\%$\downarrow$} & \multicolumn{1}{c|}{0.4\%$\downarrow$} & \multicolumn{1}{c|}{0.6\%$\downarrow$} & 0.8\%$\downarrow$ & \multicolumn{1}{c|}{0.2\%$\downarrow$} & \multicolumn{1}{c|}{0.4\%$\downarrow$} & \multicolumn{1}{c|}{0.6\%$\downarrow$} & 0.8\%$\downarrow$ & \multicolumn{1}{c|}{0.2\%$\downarrow$} & \multicolumn{1}{c|}{0.4\%$\downarrow$} & \multicolumn{1}{c|}{0.6\%$\downarrow$} & 0.8\%$\downarrow$ \\ \hline
\multicolumn{1}{c|}{\multirow{3}{*}{\begin{tabular}[c]{@{}c@{}}Fixed\\ Ratio\end{tabular}}}     & \multicolumn{2}{c|}{PU-Net}  & \Checkmark   & \multicolumn{1}{c|}{13.50}                             & \multicolumn{1}{c|}{13.10}                             & \multicolumn{1}{c|}{13.05}                             & 12.98                             & \multicolumn{1}{c|}{13.12}                             & \multicolumn{1}{c|}{12.99}                             & \multicolumn{1}{c|}{\underline{12.90}}                             & \underline{12.84}                             & \multicolumn{1}{c|}{13.20}                             & \multicolumn{1}{c|}{13.02}                             & \multicolumn{1}{c|}{12.99}                             & 12.94                             \\ \cline{2-16} 
\multicolumn{1}{c|}{}                                                                           & \multicolumn{2}{c|}{PU-GAN}  & \Checkmark   & \multicolumn{1}{c|}{13.50}                             & \multicolumn{1}{c|}{13.19}                             & \multicolumn{1}{c|}{13.11}                             & 13.06                             & \multicolumn{1}{c|}{\underline{13.05}}                             & \multicolumn{1}{c|}{12.95}                             & \multicolumn{1}{c|}{\underline{12.90}}                             & 12.88                             & \multicolumn{1}{c|}{13.13}                             & \multicolumn{1}{c|}{\underline{12.98}}                             & \multicolumn{1}{c|}{\underline{12.94}}                             & \underline{12.89}                             \\ \cline{2-16} 
\multicolumn{1}{c|}{}                                                                           & \multicolumn{2}{c|}{PU-GCN}  & \Checkmark   & \multicolumn{1}{c|}{13.34}                             & \multicolumn{1}{c|}{13.09}                             & \multicolumn{1}{c|}{12.99}                             & 12.95                             & \multicolumn{1}{c|}{13.15}                             & \multicolumn{1}{c|}{13.01}                             & \multicolumn{1}{c|}{12.95}                             & 12.95                             & \multicolumn{1}{c|}{\bf13.01}                             & \multicolumn{1}{c|}{\bf12.86}                             & \multicolumn{1}{c|}{\bf12.86}                             & \bf12.85                             \\ \hline
\multicolumn{1}{c|}{\multirow{4}{*}{\begin{tabular}[c]{@{}c@{}}Arbitrary\\ Ratio\end{tabular}}} & \multicolumn{2}{c|}{Grad-PU} & \Checkmark   & \multicolumn{1}{c|}{13.40}                             & \multicolumn{1}{c|}{12.98}                             & \multicolumn{1}{c|}{\underline{12.94}}                             & \underline{12.88}                             & \multicolumn{1}{c|}{13.18}                             & \multicolumn{1}{c|}{13.00}                             & \multicolumn{1}{c|}{12.97}                             & 12.95                             & \multicolumn{1}{c|}{13.25}                             & \multicolumn{1}{c|}{13.05}                             & \multicolumn{1}{c|}{12.99}                             & 12.95                             \\ \cline{2-16} 
\multicolumn{1}{c|}{}                                                                           & \multicolumn{2}{c|}{SAPCU}   & \XSolidBrush & \multicolumn{1}{c|}{13.30}                             & \multicolumn{1}{c|}{13.01}                             & \multicolumn{1}{c|}{\underline{12.94}}                             & 12.89                             & \multicolumn{1}{c|}{\underline{13.05}}                             & \multicolumn{1}{c|}{\underline{12.90}}                             & \multicolumn{1}{c|}{\bf12.85}                             & \bf12.81                             & \multicolumn{1}{c|}{13.10}                             & \multicolumn{1}{c|}{13.01}                             & \multicolumn{1}{c|}{12.96}                             & 12.91                             \\ \cline{2-16} 
\multicolumn{1}{c|}{}                                                                           & \multicolumn{2}{c|}{PUSS-AS} & \XSolidBrush & \multicolumn{1}{c|}{\underline{13.29}}                             & \multicolumn{1}{c|}{\bf12.93}                             & \multicolumn{1}{c|}{\underline{12.94}}                             & \bf12.86                             & \multicolumn{1}{c|}{13.07}                             & \multicolumn{1}{c|}{12.96}                             & \multicolumn{1}{c|}{12.91}                             & 12.86                             & \multicolumn{1}{c|}{13.14}                             & \multicolumn{1}{c|}{13.03}                             & \multicolumn{1}{c|}{13.00}                             & 12.97                             \\ \cline{2-16} 
\multicolumn{1}{c|}{}                                                                           & \multicolumn{2}{c|}{Ours}    & \XSolidBrush & \multicolumn{1}{c|}{\bf13.20}           & \multicolumn{1}{c|}{\underline{12.94}}           & \multicolumn{1}{c|}{\bf12.92}           & 12.89           & \multicolumn{1}{c|}{\bf13.02}           & \multicolumn{1}{c|}{\bf12.89}           & \multicolumn{1}{c|}{\underline{12.90}}           & 12.87           & \multicolumn{1}{c|}{\underline{13.06}}           & \multicolumn{1}{c|}{13.01}           & \multicolumn{1}{c|}{12.98}           & 12.95           \\ \hline
\multicolumn{4}{c|}{Ranking(S/A/AL)}                                                                                                                         & \multicolumn{1}{c|}{1/1/1}                             & \multicolumn{1}{c|}{2/2/2}                             & \multicolumn{1}{c|}{1/1/1}                             & 2/3/3                             & \multicolumn{1}{c|}{1/1/1}                             & \multicolumn{1}{c|}{1/1/1}                             & \multicolumn{1}{c|}{2/2/2}                             & 3/3/4                             & \multicolumn{1}{c|}{1/1/2}                             & \multicolumn{1}{c|}{1/1/3}                             & \multicolumn{1}{c|}{2/2/4}                             & 2/2/5                           \\ \hline
\end{tabular}
%}
\caption{Uniformity performance in terms of NUC scores. %`F', `S', `A' and `AL' have the same meaning as Table \ref{tab:tab_1}.
}
\label{tab:tab_2}
\end{table*}

\begin{figure*}[t]
    \centering
    \includegraphics[width=0.9\linewidth]{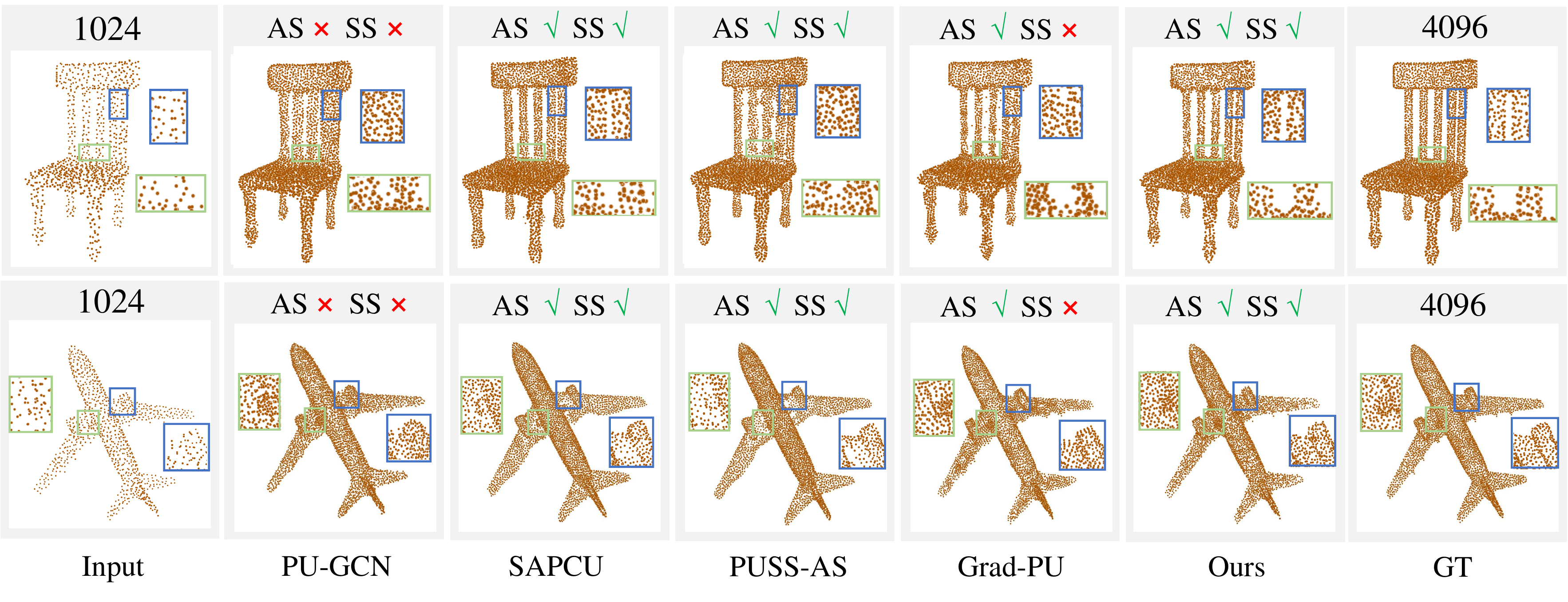}
    \caption{$4 \times$ point cloud upsampling results of \textit{chair} and \textit{airplane}, with input size of $1024$. `AS' indicates arbitrary-scale methods and `SS' indicates self-supervised methods. Please zoom in for better viewing.
}
    \label{fig:fig_3}
\end{figure*}

\begin{figure}[t]
    \centering
    \includegraphics[width=0.80\linewidth]{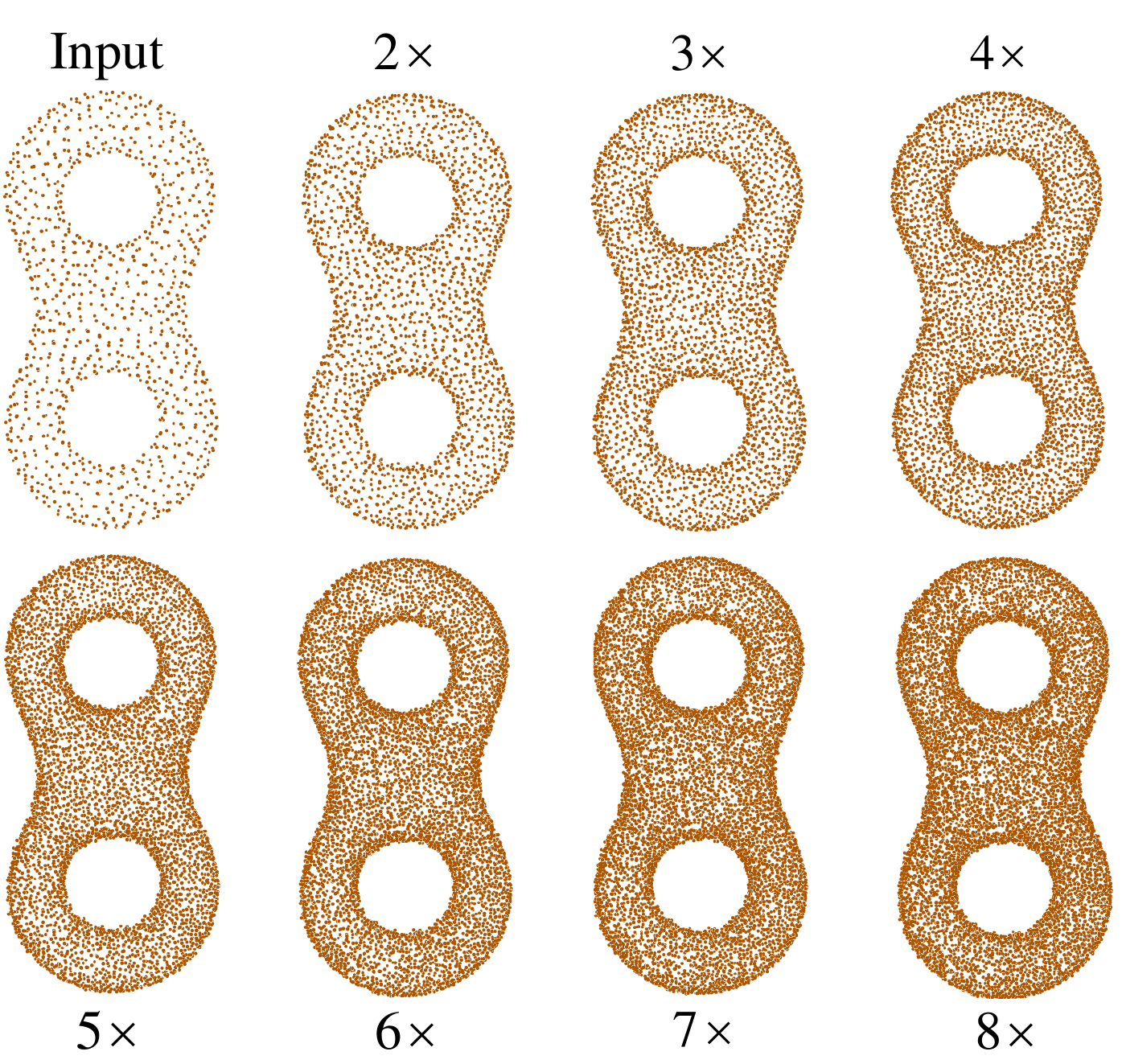}
    \caption{Results with different upsampling ratios of \textit{eight}.
}
    \label{fig:fig_4}
\end{figure}

\begin{figure}[t]
    \centering
    \includegraphics[width=0.9\linewidth]{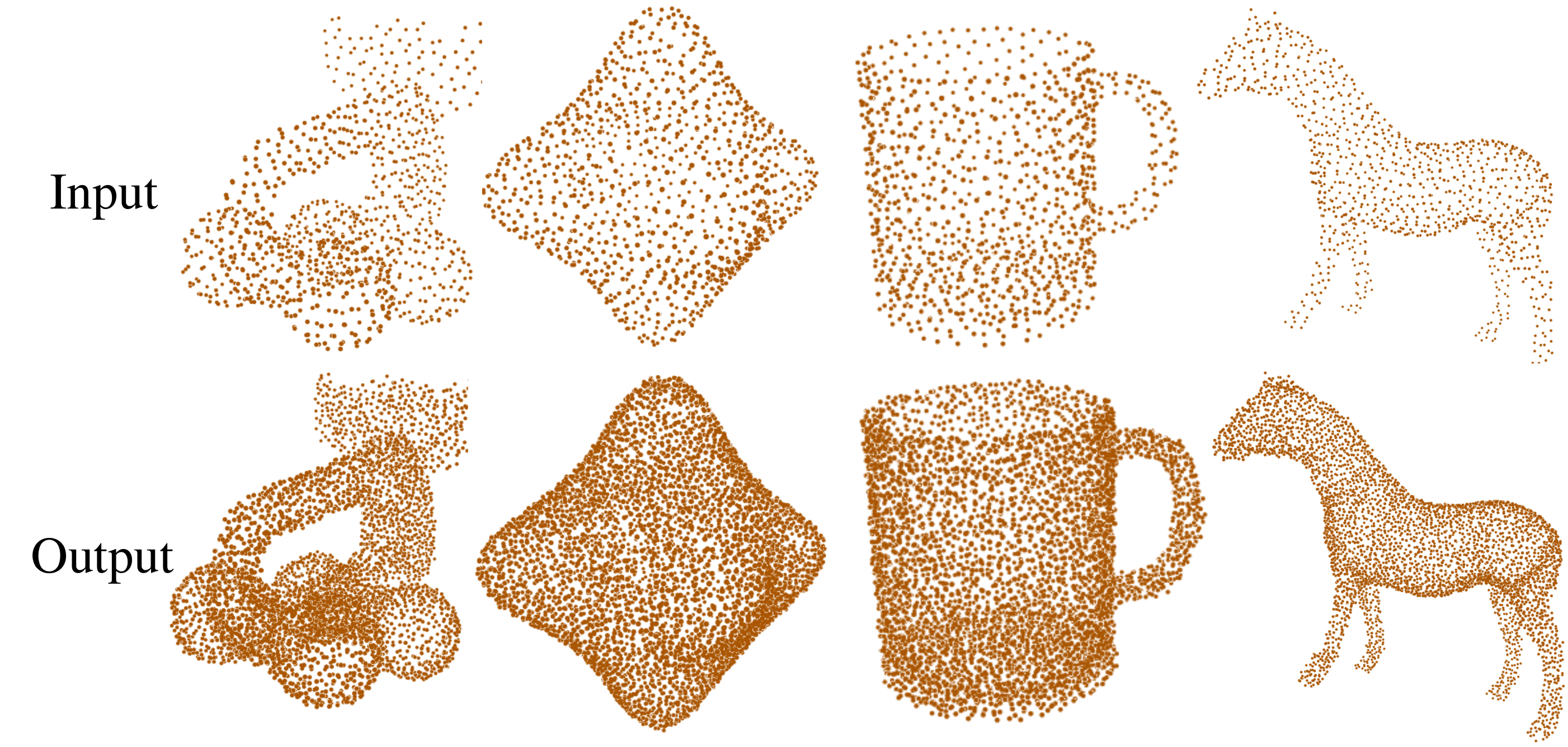}
    \caption{$4\times$ results with different unseen models.
}
    \label{fig:fig_6}
\end{figure}

\begin{figure*}[t]
    \centering
    \includegraphics[width=0.9\linewidth]{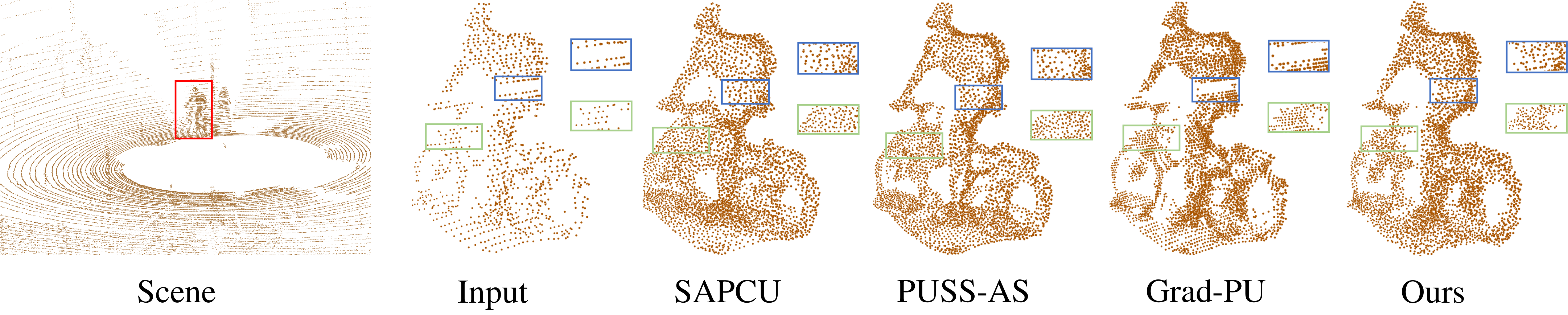}
    \caption{Point cloud upsampling results on the KITTI dataset.
}
    \label{fig:fig_5}
\end{figure*}

\subsection{Test-time Arbitrary-scale Upsampling}
Through an iterative mask-recovery operation, the completed patches are limited, making it difficult to cover all point cloud patches. Essentially, as the input mask sequence changes, the local topology structure to be restored will also change. Experiments have shown that the same point cloud patch can restore the same local topology under different mask sequences. At the same time, there are differences in the internal distribution of the point clouds.

Based on this discovery, we propose the MMR module, which utilizes multiple mask recovery operations to generate non-repeating upsampling points. This approach enables arbitrary-scale upsampling at testing time. As shown in Fig. \ref{fig:fig_1}, first, we partition the input point cloud $\mathcal{P}_{in}\in \mathbb{R}^{N\times 3}$ into patches $\mathcal{P}_{patch} \in \mathbb{R}{^{n \times k \times 3}}$. Second, we import different mask sequences $\{S_1, S_2,..., S_q\}$ into the trained model, where $S_j$ is a $0-1$ sequence of length $n$. The mask object uniformly traverses all point cloud patches. Various deformed point cloud patches are obtained through all mask sequences: 
\begin{equation}
    \{P_{S_1}, P_{S_2},..., P_{S_q}\}  \,, P_{S_j} \in \mathbb{R}{^{mnk \times 3}}\,.
    \label{formula_10}
\end{equation}
Third, we merge all of them with $\mathcal{P}_{in}$ to get a dense point cloud. Finally, we remove a small number of outliers through a simple outlier removal module \cite{zhao2022self} and use FPS to obtain the desired upsampled point cloud $\mathcal{P}_{out}\in \mathbb{R}^{rN\times 3}$, where $r$ is the upsampling ratio. Please refer to the supplementary materials for more details.

\section{Experiment}
\subsection{Comparative Study}
We compare our SPU-IMR with several state-of-the-art (SoTA) methods. They can be categorized into fixed-ratio upsampling methods: PU-Net \cite{yu2018pu}, MPU \cite{yifan2019patch}, PU-GAN \cite{li2019pu}, PU-GCN \cite{qian2021pu}, which are trained and tested with different upsampling ratios. Arbitrary ratio upsampling methods: Grad-PU \cite{he2023grad}, SAPCU \cite{zhao2022self}, and PUSS-AS \cite{zhao2023self}, which are trained only once. Notably, SAPCU \cite{zhao2022self} and PUSS-AS \cite{zhao2023self} are self-supervised upsampling methods, while others are fully supervised methods. All these methods are trained and tested using the code provided in their papers.

We employ the dataset ~\cite{zhao2022self, zhao2023self} used for training and testing. It consists of $4719$ point clouds split into $3672$ training samples and $1047$ testing samples. Each sample contains 2048 points, and we further down-sample to 1024 points through FPS to form the test set and the training set. It is noteworthy that the test samples \cite{zhao2022self, zhao2023self} used are from the dataset adopted by \cite{yu2018pu, ye2021meta}.

\subsection{Implement Details.}
\subsubsection{Parameters Setting.}
During the training phase, each sparse point cloud ($N=1024$) is partitioned into 256 patches, and each patch contains 16 points. The mask ratio $m$ is assigned to 0.6. The number of iteration layers $L$ is set to 3. As for the testing phase, we assign $q=8$ mask sequences. These mask sequences are generated randomly. Please refer to the supplementary materials for more details.
%The coefficient of the outlier removal module is set as $2.0$. 

\subsubsection{Training Details.}
The training process is conducted on a server with four GeForce RTX 3090 GPUs. The networks are trained for 6000 epochs with a batch size of 128. For training details, we use an AdamW optimizer \cite{loshchilov2017decoupled} and cosine learning rate decay \cite{loshchilov2016sgdr}. The initial learning rate is set to 0.001, with a weight decay of 0.05.

\subsection{Quantitative Analysis}
\subsubsection{Evaluation Metrics.}
To comprehensively evaluate the upsampled point clouds, we utilize several popular metrics: Chamfer Distance (CD), which measures the similarity between two point clouds. EMD, which measures the minimum cost of turning one of the point cloud into the other. Hausdorff Distance (HD), which effectively evaluates the differences in outliers between point clouds. F-score \cite{wu2019point}, which treats upsampling as a classification problem, where a higher F-score is achieved for more accurate point clouds. Point-to-Face distance (P2F) \cite{yu2018pu}, which evaluates the average distance between the predicted generated point clouds and the ground truth meshes. Normalized Uniformity Coefficient (NUC) \cite{yu2018pu}, which evaluates the uniformity of points on randomly selected disks with different area percentages $p=0.2\%, 0.4\%, 0.6\%, 0.8\%$. 

\subsubsection{Quantitative Evaluation.}
As illustrated in Table \ref{tab:tab_1}, we compare our method with other SoTA approaches at three upsampling ratios $\left[ {2 \times,4 \times,8 \times } \right]$ in terms of CD, EMD, HD, F-Score, P2F. Even though we are trained in a self-supervised manner, our method achieves competitive results with other SoTA upsampling methods. In terms of the $2 \times$, our method performs best on EMD, HD, and F-Score. In terms of the $4 \times$, we achieve the best results on P2F. Moreover, CD and HD are only lower than the best fully supervised method. In terms of the $8 \times$, we get rank \#1 on F-Score and P2F. The performance results indicate that our proposed method can generate highly accurate upsampled point clouds with minimal errors.

As illustrated in Table \ref{tab:tab_2}, we compare our method with other approaches at upsampling scales of $\left[ {2 \times,4 \times,8 \times } \right]$ in terms of NUC. Our NUC score in most situations is higher than other methods, indicating that our method produces more uniformly distributed point cloud results. Using different mask sequences, we can generate an arbitrary number of points in each point cloud patch, resulting in a relatively uniform distribution of points after FPS.

\subsection{Qualitative Evaluation}
\subsubsection{Visual Comparison.}
As shown in Fig. \ref{fig:fig_3}, we compare the visualization results of different upsampling methods at a $4\times$ scale on $chair$ and $airplane$. %The input contains $1024$ points.

For $chair$, most SoTA methods generate noise between two pillars. While SAPCU avoids this additional noise, it does have a gap at the connection between the seat and the pillar and generates some noticeable noise directly beneath the chair. By contrast, our method not only eliminates significant noise but also better restores the chair's local structure. For $airplane$, most methods generate sparse or unevenly distributed upsampling point clouds at the junction between the wing and the fuselage. Moreover, PUGCN does not accurately reproduce the detailed structure of the jet nozzle. Our method maintains the uniform distribution of the input point cloud and closely aligns with the ground truth.

\subsubsection{Results on PU1K under Different Upsampling Ratios.}
    PU1K dataset \cite{qian2021pu} consists of 1147 samples. To evaluate the generalization capability of our method, we test our model under cross-category samples. The input contains 1024 points.

As shown in Fig. \ref{fig:fig_4}, on unseen sample \textit{eight}, our method is capable of preserving the details of objects well across different upsampling ratios. Results in Fig. \ref{fig:fig_6} demonstrate our method's good generalization across various categories.

\subsubsection{Results on the Real-world KITTI Dataset.}
The KITTI dataset \cite{geiger2013vision} is a benchmark for autonomous driving and computer vision. It includes diverse real-world driving scenarios captured with LiDAR. To further test the practical application value of our model, we evaluated it on the real-scanned dataset KITTI. 

Fig. \ref{fig:fig_5} illustrates the comparison between our method and SoTA arbitrary ratio upsampling methods. Both SAPCU and PUSS-AS completely cover the front wheels of the car, which is unreasonable. Moreover, Grad-PU shows a clear blank area around the waist of the human and the area where the handlebars are held. These results highlight their weaknesses in dealing with noise and accurately improving local structures, whereas our proposed method performs well.

\subsection{Analysis Experiments}
\subsubsection{Varying Mask Ratios.}
We test the performance of the upsampled point cloud with different mask ratios ranging from 0.3 to 0.8. Other training and testing processes are under the same conditions. Taking $4\times$ as an example, we still use CD, EMD, HD, F-Score and P2F to evaluate these upsampled point clouds comprehensively. 

As shown in Table \ref{tab:tab_3}, at lower mask ratios, such as 0.3 and 0.4, the evaluation results are significantly different from others. As the mask ratio increases to 0.5 and 0.6, all indicators perform well. However, when the mask ratio further increases to 0.7 and 0.8, the performance of most metrics slightly decreases. We believe that with a small mask ratio, the recovered masked point cloud patches, derived from abundant known information, tend to overfit. This causes the points within each generated point cloud patch to cluster together, leading to abnormal evaluation outcomes. Conversely, with a large mask ratio, the recovery results are suboptimal due to the excessive amount of missing information. Based on the above analysis, we choose $m=0.6$. 

\subsubsection{Varying Iterations.}
We evaluate the model's performance across different iteration times. Specifically, we set the iterative layer $L$ to $\left[ {1,2,3,4,5} \right]$ respectively and assess their performance on CD, EMD, HD, F-Score, and P2F. All other training and testing processes remain consistent.

As shown in Table \ref{tab:tab_4}, the model's performance improves with iteration layers increasing. When $L>3$, the performance shows no significant improvement. Therefore, to conserve computational resources, we choose $L=3$.

\begin{table}[t]
\centering
\setlength{\tabcolsep}{1.8mm}
\begin{tabular}{c|c|c|c|c|c}
\hline
\multirow{2}{*}{\begin{tabular}[c]{@{}c@{}}Varying \\ mask ratios\end{tabular}} & \multirow{2}{*}{CD$\downarrow$} & \multirow{2}{*}{EMD$\downarrow$} & \multirow{2}{*}{HD$\downarrow$} & \multirow{2}{*}{F-Score$\uparrow$} & \multirow{2}{*}{P2F$\downarrow$} \\
                                                                                &                                 &                                  &                                 &                                    &                                  \\ \hline
0.3                                                                             & 14.7                            & 48.5                             & 49.7                            & 49.6                               & 3.01                             \\ \hline
0.4                                                                             & 15.3                            & 36.7                             & 40.3                            & 48.8                               & 5.08                             \\ \hline
0.5                                                                             & 14.1                            & 24.9                             & 36.7                            & 49.1                               & 6.69                             \\ \hline
0.6                                                                             & 13.9                            & 22.2                             & 22.7                            & 49.2                               & 6.85                             \\ \hline
0.7                                                                             & 14.5                            & 23.8                             & 25.4                            & 49.1                               & 6.95                             \\ \hline
0.8                                                                             & 14.3                            & 23.6                             & 25.5                            & 49.2                               & 7.14                             \\ \hline
\end{tabular}
%}
\caption{Results with varying mask ratios.}
\label{tab:tab_3}
\end{table}

\begin{table}[t]
\centering
\begin{tabular}{c|c|c|c|c|c}
\hline
\multirow{2}{*}{\begin{tabular}[c]{@{}c@{}}Varying\\ iterations\end{tabular}} & \multirow{2}{*}{CD$\downarrow$} & \multirow{2}{*}{EMD$\downarrow$} & \multirow{2}{*}{HD$\downarrow$} & \multirow{2}{*}{F-Score$\uparrow$} & \multirow{2}{*}{P2F$\downarrow$} \\
                                                                              &                                 &                                  &                                 &                                    &                                  \\ \hline
1                                                                             & 15.9                            & 33.1                             & 34.8                            & 48.7                               & 6.38                             \\ \hline
2                                                                             & 14.0                            & 23.2                             & 23.7                            & 49.2                               & 6.84                             \\ \hline
3                                                                             & 13.9                            & 22.2                             & 22.7                            & 49.2                               & 6.85                             \\ \hline
4                                                                             & 14.5                            & 23.8                             & 25.4                            & 49.1                               & 6.95                             \\ \hline
5                                                                             & 14.3                            & 23.6                             & 25.5                            & 49.2                               & 7.14                             \\ \hline
\end{tabular}
%}
\caption{Results with varying iterations.}
\label{tab:tab_4}
\end{table}

\section{Conclusion}
In this paper, we proposed a novel iterative mask-recovery framework based on self-supervised learning. Our method allows for arbitrary-scale upsampling with only one training process. Extensive experimental results demonstrate that our method can produce high-quality dense point clouds that are uniform and complete, and achieve competitive objective performance and even better visual performance compared with existing methods.

\section{Acknowledgments}
This work is supported in part by the National Natural Science Foundation of China (No. 62372377) and Guangdong Basic and Applied Basic Research Foundation (2023B1515120026).

\bibliography{aaai25}

\end{document}